 \documentclass[final,5p,times,twocolumn,authoryear]{elsarticle}

\usepackage{amssymb}
\usepackage{lipsum}
\usepackage{mhchem}							

\newcommand{\bea}{\begin{eqnarray}}
\newcommand{\eea}{\end{eqnarray}}
\newcommand{\lp}{\left(}
\newcommand{\rp}{\right)}
\newcommand{\lb}{\left[}
\newcommand{\rb}{\right]}
\newcommand{\non}{\nonumber}
\newcommand{\dopr}{\prime\prime}
\newcommand{\rv}{{\bf r}}
\newcommand{\rvp}{\rv^{\prime}}
\newcommand{\xv}{{\bf x}}
\newcommand{\xvp}{\xv^{\prime}}
\newcommand{\rpr}{r^{\prime}}
\newcommand{\nal}{n_{\alpha}}
\newcommand{\nnal}{N_{\alpha}}
\newcommand{\nnp}{N_{p}}
\newcommand{\nnn}{N_{n}}
\newcommand{\rrz}{R_{0}}
\newcommand{\ovzer}{\overline{v}_{0}}
\newcommand{\intd}{\int\displaylimits}	
\newcommand{\sumd}{\sum\displaylimits}
\newcommand{\ggamdir}{\Gamma^{\lp\textrm{dir}\rp}}
\newcommand{\ggamexc}{\Gamma^{\lp\textrm{exc}\rp}}
\newcommand{\psiam}{\psi_{am}}
\newcommand{\epsa}{\epsilon_{a}}

\journal{Physics Letters B}

\begin{document}

\begin{frontmatter}



\title{Emission processes in a self--consistent field}


\author[add1]{A. Dumitrescu}
\author[add1,add2]{D.S. Delion}
\affiliation[add1]{organization={Horia Hulubei National Institute for 
                   R\&D in Physics and Nuclear Engineering},           
            addressline={No. 30 Reactorului Street}, 
            city={Magurele},
            postcode={077125}, 
            state={Ilfov},
            country={Romania}}
\affiliation[add2]{organization={Academy of Romanian Scientists},
            addressline={No. 3 Ilfov Street},
            city={Bucharest},
            postcode={050044},
            state={sector 5},
            country={Romania}}

\begin{abstract}
We present a microscopic description of cluster emission processes 
within the Cluster--Hartree--Fock (CHF) self--consistent field (SCF) 
theory. The starting point is a Woods--Saxon (WS) mean field (MF) with 
spin--orbit and Coulomb terms. Pairing is treated through standard 
Bardeen--Cooper--Schrieffer (BCS) quasiparticles. The residual two--body
interaction is given by a density--dependent Wigner force having a 
Gaussian shape with a center of mass (com) correction located in a
region of low nuclear density slightly beyond the geometrical contact 
radius of a system comprised from a nucleus and a surface cluster. We 
show that such a description adequately reproduces the ground state (gs)
shape of a spherical nucleus while the surface correction enhances the 
radial tail of single particle (sp) orbitals, thus allowing for an 
adequate description of the $\alpha$-decay width for unstable systems.
\end{abstract}




\end{frontmatter}




\section{Introduction}
\label{introduction}

The phenomenon of radioactivity was discovered by Becquerel in 1896. It 
was largely ellucidated in the following years by Rutherford and Soddy 
as the $\alpha$, $\beta$ and $\gamma$ classification of what was to be 
recognized as nuclear radiation \citep{Rad17} following the discovery of 
the atomic nucleus through $\alpha$--scattering experiments by 
\cite{Rut11}. The first empirical formula relating the decay constant to 
the energy of the emitted $\alpha$--particles was the law of 
\cite{Gei11}. In the following century, the phenomenon of 
$\alpha$--decay became a fundamental probe of nuclear structure through 
the medium of the strong nuclear force. It revealed evidence for the 
existence of low--lying nuclear excitations \citep{Ros29} and the 
$\alpha$--clustered states of light nuclei \citep{Oer06,Fre18}. It was 
also a critical tool for the detection of superheavy nuclei in the 
ongoing pursuit of an island of stability, the boundaries of the chart 
of nuclides and the limit of the periodic table \citep{Smi24}. The 
synthesis of atomic nuclei beyond the lightest elements begins in stars 
through the triple $\alpha$---process where the structure of the Hoyle 
state of C$^{12}$ plays a crucial role \citep{Toh17}. Recently, 
knock--out reactions in neutron--rich Sn isotopes have shown empirically 
that $\alpha$--particles are located on the surface of atomic nuclei, 
with potentially fundamental implications for complex nucleosynthesis as
in the r--process or neutron star physics \citep{Tan21}.

The first theories of $\alpha$--decay seeking to explain the 
Geiger--Nuttall law were formulated by \cite{Gam28} and independently 
by \cite{Gur28} using the wave mechanics newly developed at the time. 
Thus, $\alpha$--decay was the first phenomenon to be understood as a 
tunneling process on the basis of the probabilistic interpretation of 
quantum mechanics. Later, \cite{Gam49} found a classical analogy with 
wave optics and the transmission of light through a thin reflective 
layer. Major formal developments on quantum tunneling followed from the 
formulation of the R--matrix theory of nuclear reactions by 
\cite{Lan58}. 

The modern theoretical problem that remains open regards the formation 
of clusters on the nuclear surface, i.e. the calculation of the overlap 
between the many--particle gs configuration of the parent nucleus and a 
configuration consisting of a daughter nucleus and surface cluster. This 
is a complicated many--body effect involving significant contributions 
from the continuum part of the sp nuclear spectrum and as such is beyond 
the reach of MF approximations. It has been studied extensively within 
the shell model \citep{Oko12}, but even when including a very large 
number of configurations in the calculations, the absolute decay widths 
still differ from experimental values by at least one order of magnitude 
\citep{Bet12}. 

The phenomenological solution to this problem consists in describing the
decay process through the motion of a cluster in an attractive  
potential located on the nuclear surface. In the framework of R--matrix 
theory applied to $\alpha$--decay, this approach predicts a linear 
dependence between the logarithm of the reduced decay width (given by 
the ratio of the decay width to the Coulomb penetrability) and the 
fragmentation potential defined as the difference between the 
Coulomb barrier and decay Q--value, as shown by \cite{Del09}. 
This was later proved to remain true for many strong emission processes,
from proton radioactivity to the emission of heavier clusters 
\citep{Dum22}. 

A very popular many--body method with vast applicability
in nuclear structure is density functional theory 
\citep{Lal04,Dob11,Col20,Col22}. It allows for the investigation of
highly exotic modes at the nuclear and astrophysical level 
\citep{Paa07}, can treat complex dynamical phenomena such as nuclear 
fission \citep{Sch16} and offers a relativistic perspective on cluster
states \citep{Ebr14}. Other approaches to the clustering problem make 
use of nonlinear dynamics and the motion of solitons on quantum 
droplets \citep{Car21}.

In this letter we expand on several previous works \citep{Del13,Dum23} 
to present a self--consistent calculation for the gs of an 
$\alpha$--decaying nucleus in terms of proton (p) and neutron (n) 
degrees of freedom in a HF field developed from a density--dependent 
Gaussian--shaped residual interaction enhanced on the nuclear surface. 
Section \ref{theory} details  the formal development of the theory, 
Section \ref{numap} provides an illustration of the method through 
numerical application and Section \ref{conc} presents the conclusions.  

\section{CHF theory}
\label{theory}

In this subsection we extend the standard HF scheme to a cluster--HF 
procedure, illustrating the cluster by an $\alpha$-particle. The HF 
approximation is a well-established technique for the calculation of gs 
properties for many--body systems with mutual interactions between 
constituents. A number of comprehensive reviews of the method pertaining 
to the atomic nucleus have been written \citep{Que78,Gog86} and some 
computer codes are available \citep{Col13}. Here we follow an original 
approach based on a residual interaction enhanced on the nuclear 
surface. The properties of the interaction are developed in Subsection 
\ref{resv}, the relevant CHF equations are derived in Subsection 
\ref{hfeq} and the $\alpha$--cluster amplitude is calculated in 
Subsection \ref{forma}.
 
\subsection{Residual interaction}
\label{resv}

At low energies, the nucleon--nucleon interactions are mainly attractive
\citep{Rin80}. The most important part is the central force, which is 
decomposed into four terms. When one neglects the spin-- and 
isospin--dependent parts, which are quite relevant at higher energies 
but not overly important for the cluster emission process to be 
discussed here, one retains the Wigner term defined in coordinate space. 
We choose for it a surface Gaussian--shaped two--body interaction (SGI) 
of the form
\bea
\label{resin}
v\lp\rv,\rvp\rp=-\ovzer\kappa\lp\rpr\rp 
e^{-\frac{\vert\rv-\rvp\vert^{2}}{b^{2}}}
\lb 1+x_{c}e^{-\frac{\lp R-\rrz\rp^{2}}{B^{2}}}\rb.
\eea
Here, $\ovzer$ is the interaction constant and $\rv$ and $\rvp$ are the
radial vectors of any two particles. The two exponentials decompose
the interaction in a relative part at the radius $r=\vert\rv-\rvp\vert$ 
and com part at the radius $R=\vert\rv+\rvp\vert/2$ centered on the 
surface of a sphere of radius $\rrz$. The two components have effective 
lengths $b$ and $B$ respectively and $x_{c}$ is a control parameter that
enhances the surface term. $\kappa\lp\rpr\rp$ is an effective 
density--dependent term
\bea
\kappa\lp\rpr\rp=\frac{\rho^{\lp 0\rp}\lp\rpr\rp}
{\langle\rho\lp\rpr\rp\rangle}
\eea
where $\rho^{\lp 0\rp}\lp\rpr\rp$ is the initial density and
$\langle\rho\lp\rpr\rp\rangle$ is the local average of the density in 
a small region around every point during a given iteration. Such a 
density--dependence takes into account the screening of the interaction 
due to neighboring nucleons. Moreover, it stabilizes the HF iterative 
procedure by preventing pathological behaviors in the central region of 
the nucleus. The surface term in Eq. (\ref{resin}) provides a 
phenomenological description of nucleonic clustering at low densities, 
beyond the Mott transition point \citep{Rop98}. Its use also avoids the 
rather complicated cranking procedure imposing a given quadrupole moment
within the Hartree-Fock--Bogoliubov (HFB) method which relies on the 
generalized quasiparticle transformation.

It is to be understood that the parameters $\ovzer, b, \rrz, B$ and 
$x_{c}$ are in principle isospin--dependent and may have different 
values for protons and neutrons. The multipole expansion of the 
interaction is relevant for the calculation of the SCF, the multipole of 
order $L$ being defined as
\bea
v_{L}\lp r,\rpr\rp=\intd_{-1}^{1}\mathrm{d}\gamma^{\prime}
\zeta\lp\rv,\rvp\rp\mathcal{P}_{L}\lp\gamma^{\prime}\rp
\eea
where $\mathcal{P}_{L}$ is a Legendre polynomial, $\gamma^{\prime}$ is 
the cosine of the angle between the particle vectors and
$\zeta\lp\rv,\rvp\rp=-\frac{1}{\ovzer}v\lp\rv,\rvp\rp$. 
The matrix elements of $\zeta\lp\rv,\rvp\rp$ are calculated through the 
standard methods of the Talmi--Moshinsky (TM) transformation.

\subsection{CHF equations}
\label{hfeq}

For the proton and neutron fields having spatial and spin degrees of 
freedom $\xv$, the CHF equations read \citep{Rin80}
\bea\label{hfbcs}
&~&\lb -\frac{\hbar^{2}}{2\mu}{\bf\nabla}^{2}+\ggamdir\lp\rv\rp\rb
\psiam\lp\xv\rp\non\\
&+&\int\mathrm{d}^{3}\rvp\ggamexc\lp\rv,\rvp\rp
\psiam\lp\xvp\rp=\epsa\psiam\lp\xv\rp
\eea
where $\mu$ is the nucleon mass and the eigenvalue index $a$ is a 
shorthand for the set of quantum numbers containing the isospin $\tau$, 
sp energy $\epsilon$, orbital angular momentum $\ell$ and total 
angular momentum $j$. $m$ is the total angular momentum projection along 
the $z$ axis. The direct and exchange terms are evaluated by folding the
residual interaction over nucleon densities
\bea
\ggamdir\lp\rv\rp&=&\int\mathrm{d}^{3}\rv v\lp\rv,\rvp\rp\rho\lp\rvp\rp
\non\\
\ggamexc\lp\rv,\rvp\rp&=&-\int\mathrm{d}^{3}\rv v\lp\rv,\rvp\rp
\rho\lp\rv,\rvp\rp
\eea
with the densities in turn being expanded in the basis of the sp wave 
functions
\bea
\rho\lp\rvp\rp&=&\sumd_{c}v_{c}^{2}\sumd_{s}
\psi_{cs}^{\dagger}\lp\xvp\rp\psi_{cs}\lp\xvp\rp\non\\
\rho\lp\rv,\rvp\rp&=&\sumd_{c}v_{c}^{2}\sumd_{s}
\psi_{cs}^{\dagger}\lp\xvp\rp\psi_{cs}\lp\xv\rp.
\eea
The expansion coefficients $v_{c}$ are the usual BCS occupation 
amplitudes. We seek solutions of Eqs. (\ref{hfbcs}) of the form
\bea
\psiam\lp\xv\rp=\frac{1}{r}u_{a}\lp r\rp\lb i^{\ell_{a}}Y_{\ell_{a}}
\otimes\chi_{\frac{1}{2}}\rb_{j_{a}m}
\eea
where the spin--orbit function is expressed as the coupling between a 
spherical harmonic and spinor. One can then readily arrive at a set of
coupled second--order radial differential equations reminiscent of that 
obtained by \cite{Vau67}
\bea
\label{radsys}
-u_{a}^{\dopr}\lp r\rp+
\frac{\ell_{a}\lp\ell_{a}+1\rp}{r^{2}}u_{a}\lp r\rp+
\frac{2\mu}{\hbar^{2}}\lb V_{a}\lp r\rp-\epsa\rb u_{a}\lp r\rp=0
\eea
where the equivalent local potential $V_{a}$ is given by 
\bea
&~&V_{a}\lp r\rp=-\ovzer\sqrt{4\pi}\intd_{0}^{\infty}\mathrm{d}\rpr
r^{\prime 2}\rho\lp\rpr\rp v_{0}\lp r,\rpr\rp\non\\
&+&\ovzer\sumd_{c}v_{c}^{2}\frac{u_{c}\lp r\rp}{u_{a}\lp r\rp}
\lp i\rp^{\ell_{c}-\ell_{a}}\non\\
&\times&\sumd_{L}\mathcal{I}_{L}^{\lp ac\rp}\lp r\rp
\langle j_{c} \frac{1}{2}, L 0\vert j_{a} \frac{1}{2}\rangle
\langle j_{a} \frac{1}{2}, L 0\vert j_{c} \frac{1}{2}\rangle.
\eea
The integrals $\mathcal{I}_{L}^{\lp ac\rp}$ evaluate as
\bea
\mathcal{I}_{L}^{\lp ac\rp}\lp r\rp=\intd_{0}^{\infty}\mathrm{d}\rpr 
u_{a}\lp\rpr\rp u_{c}\lp\rpr\rp v_{L}\lp r,\rpr\rp
\eea
and the quantities in brakets are ordinary Clebsch--Gordan coupling 
coefficients. For spherical systems it is sufficient to consider only
the monopole contribution in the exchange term of $V_{a}\lp r\rp$. By 
expanding the radial functions $u_{a}\lp r\rp$ in the spherical harmonic
oscillator (ho) basis, one gets an eigenvalue problem
\bea
\sumd_{n^{\prime}} H_{na,n^{\prime}a}^{\lp\beta\rp} 
d_{a}^{\lp n^{\prime}\rp}=\epsa d_{a}^{\lp n\rp}.
\eea
Here, $d_{a}^{\lp n\rp}$ is the expansion coefficient of $u_{a}\lp r\rp$ 
onto the corresponding ho state of radial quantum number $n$, $\beta$ is 
the ho parameter $\beta=\frac{\mu\omega}{\hbar}$ and the  Hamiltonian 
matrix is given by
\bea
\label{eigpro}
H_{na,n^{\prime}a}^{\lp\beta\rp}&=&
\hbar\omega\lp 2n+\ell_{a}+\frac{3}{2}\rp\delta_{nn^{\prime}}
+\langle\beta n\ell_{a}\vert V_{a}\lp r\rp
\vert\beta n^{\prime}\ell_{a}\rangle\non\\
&-&\frac{\hbar\omega}{2}\langle\beta n\ell_{a}\vert\beta r^{2}
\vert\beta n^{\prime}\ell_{a}\rangle
\eea
Thus, as an alternative to numerical integration, the CHF SCF can be 
found by solving an eigenvalue problem.

\subsection{$\alpha$--particle formation amplitude}
\label{forma}

Detailed computational aspects are covered by \cite{Del10}, so here we 
only summarize the most important results. In the spherical approach 
under study, the $\alpha$--particle formation amplitude is given by the 
overlap integral depending upon the com radius of the $\alpha$-daughter 
system
\bea
\mathcal{F}_{0}\lp R\rp=
\langle\Psi_{P}\vert\Psi_{D}\Psi_{\alpha}\rangle,
\eea
where the wave functions $\Psi$ of indices $P, D$ and $\alpha$ describe
the parent and daughter nuclei and the $\alpha$--particle.
This simple approximation is valid at distances around and beyond the 
geometrical touching radius, where low values of the density imply low 
antisymmetrization effects. Just as for the SCF, the problem is most 
conveniently analyzed in a ho representation, with the formation 
amplitude following as
\bea
\mathcal{F}_{0}\lp R\rp=\sumd_{\nnal}\mathcal{W}_{\nnal}
\mathcal{R}_{\nnal 0}^{\lp 4\beta\rp}\lp R\rp=
\sumd_{\nnal}\mathcal{F}_{\nnal 0}\lp R\rp
\eea
in terms of the radial ho functions.
$\nnal$ is the ho radial quantum 
number corresponding to an $\alpha$--particle moving with angular 
momentum $L_{\alpha}=0$. The coefficients $\mathcal{W}$ are given by the
superposition
\bea
\mathcal{W}_{\nnal}&=&8\sumd_{\nal\nnp\nnn}
\mathcal{G}_{\nnp}\mathcal{G}_{\nnn}\non\\
&\times&\langle\nal, 0;\nnal, 0;0\vert\nnp,0;\nnn,0;0\rangle
\mathcal{J}_{\nal 0}^{\lp\beta\beta_{\alpha}\rp}.
\eea
The braket represents a TM recoupling coefficient that
connects $pp$ and $nn$ states to $\alpha$--particle coordinates. The
integral involved in the expansion overlaps ho sp radial states with the 
$\alpha$--particle wavefunction. The quantities $\mathcal{G}_{N_{\tau}}$
contain nucleonic degrees of freedom for given isospin
\bea
\mathcal{G}_{N_{\tau}}&=&\sumd_{n_{1}n_{2}\ell j}
\mathcal{B}_{\tau}\lp n_{1}\ell jn_{2}\ell j;0\rp\non
\\&\times&
\langle\lp\ell\ell\rp0\lp\frac{1}{2}\frac{1}{2}\rp0;0
\vert\lp\ell\frac{1}{2}\rp j\lp\ell\frac{1}{2}\rp j;0\rangle\non\\
&\times&\sumd_{n_{\tau}}\langle n_{\tau}0N_{\tau}0;
0 n_{1}\ell n_{2}\ell;0\rangle
\mathcal{J}^{\beta\beta_{\alpha}}_{n_{\tau}0}.
\eea
The braket on the second line is the $jj-LS$ recoupling coefficient. 
The nuclear structure information in terms of the BCS amplitudes and 
ho expansion coefficients is contained in the quantities
\bea
\mathcal{B}_{\tau}\lp n_{1}\ell jn_{2}\ell j;0\rp=
\frac{\hat{j}}{\sqrt{2}}u_{\tau\epsilon\ell j}v_{\tau\epsilon\ell j}
d_{\tau\epsilon\ell j}^{\lp n_{1}\rp}
d_{\tau\epsilon\ell j}^{\lp n_{2}\rp}
\eea
where $\hat{j}=\sqrt{2j+1}$. The $p-n$ interaction can be neglected due 
to the different major shells involved in the calculation \citep{Del00}.

The formation amplitude thus obtained, playing the role of the 
$\alpha$-particle internal wave function, is then matched to the 
external outgoing Coulomb wave, giving the gs decay width 
\citep{Lan58,Del10}
\bea
\Gamma_{th}(R)=\hbar v\lb\frac{R\mathcal{F}_{0}\lp R\rp}
{G_{0}\lp\chi,\rho\rp}\rb^{2}.
\eea
Here, $v=\sqrt{2E/\mu_{\alpha}}$ is the asymptotic particle velocity 
depending upon the Q-value of the $\alpha$-decay, $E$. $G_{0}$ is the 
irregular monopole Coulomb wave which practically coincides with the 
outgoing Coulomb--Hankel wave inside the barrier. It depends upon the 
Coulomb parameter $\chi=2Z_DZ_{\alpha}e^2/\hbar v$ and the reduced 
radius $\rho=\kappa R$ = $(\mu_{\alpha} v/\hbar) R$. We will show in the 
next Section that the decay width satisfies the plateau condition by 
depending weakly upon the matching radius $R$ in a region beyond the 
geometrical touching configuration
\bea
\rrz=r_{0}\lp A_{D}^{\frac{1}{3}}+4^{\frac{1}{3}}\rp,~r_{0}=1.2
~\mathrm{fm}.
\eea 

\section{Numerical application}
\label{numap}

\begin{figure}
\centering 
\includegraphics[width=0.4\textwidth]{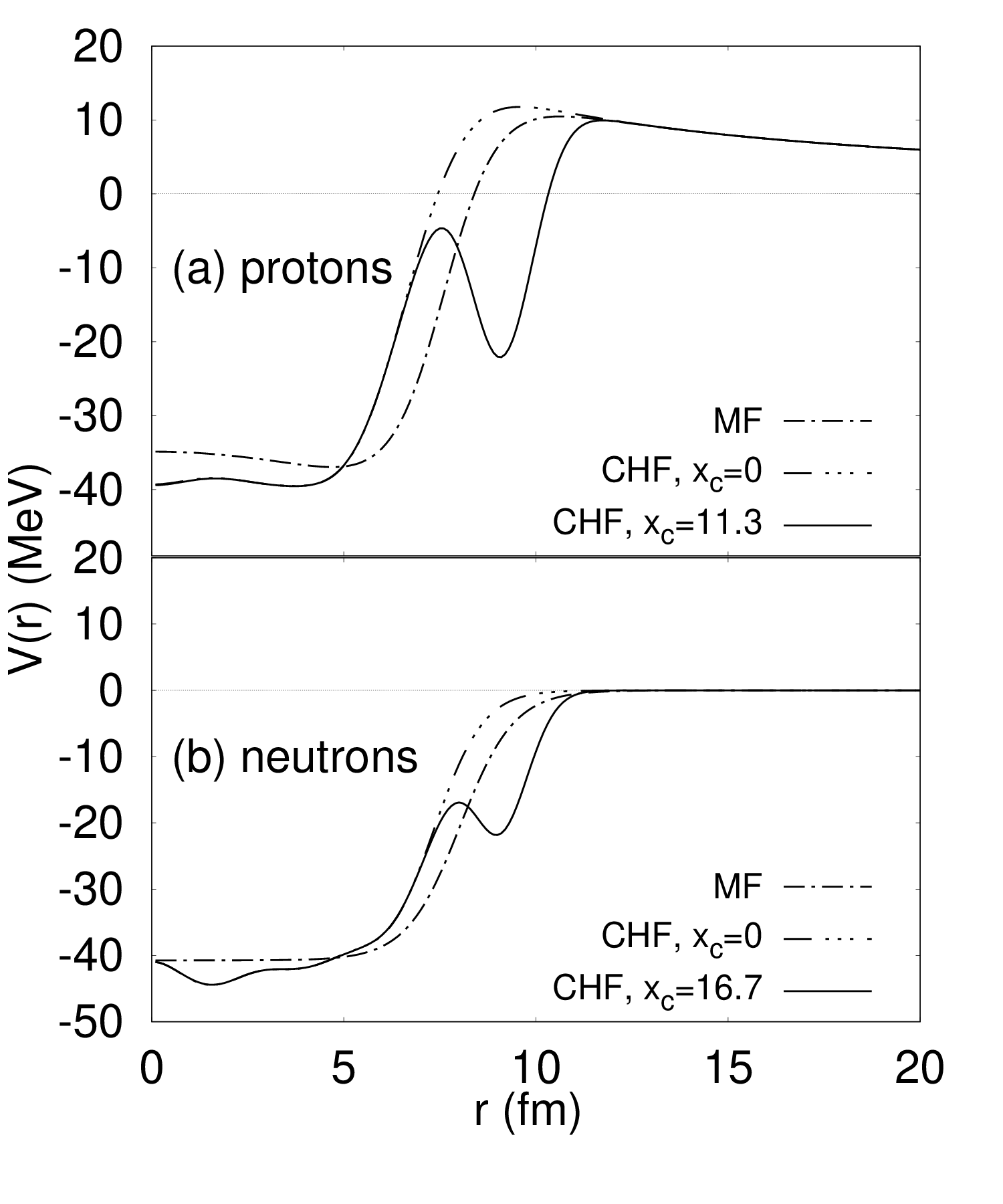}	
\caption{Direct term of the interaction potential versus radius for 
protons (a) and neutrons (b). The dot--dashed lines represent the 
initial MFs, the triple dot--dashed lines are the SCFs without the 
surface corrections and the continuous lines are the SCFs with surface 
corrections that reproduce the decay width.} 
\label{fig1}
\end{figure}

\begin{figure}
\centering 
\includegraphics[width=0.4\textwidth]{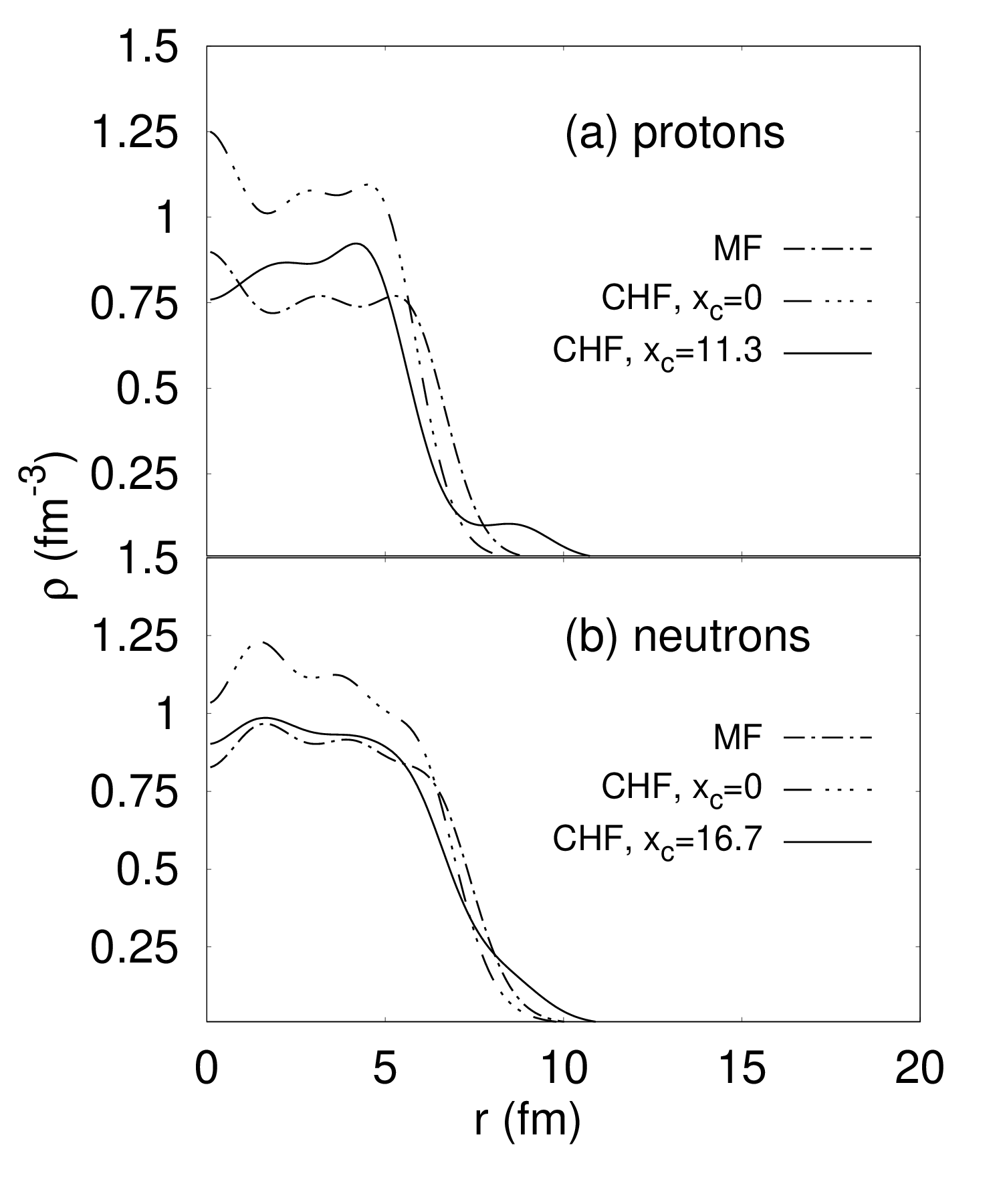}	
\caption{Density of the parent nucleus versus radius for protons (a) and 
neutrons (b). The dot--dashed lines represent the initial MF 
configurations of nucleons, the triple dot--dashed lines are the SCF 
densities without the surface corrections and the continuous lines are 
the SCF densities with surface corrections that reproduce the decay 
width.} 
\label{fig2}
\end{figure}

As an illustration of the formalism we study the decay process
$\ce{^{216}_{}Rn}\rightarrow\ce{^{212}_{}Po}+\alpha$, which has the
value of the decay width $\Gamma_{0}=1.0\cdot10^{-17}~\mathrm{MeV}~ 
\textit{1}$ \citep{Aur20}. The starting configuration is of a WS MF with 
spin--orbit and Coulomb terms in the universal parametrization of 
\cite{Dud81}. This particular parametrization was set up in the \ce{Pb}
region and is a good starting point for the calculation. It is depicted 
in Fig. \ref{fig1} by the dot--dashed lines. We use a ho basis with 
$N=12$ major shells which in this example allows for the calculation of 
wave functions up to $\approx13~\mathrm{fm}$ before they drop 
exponentially. This is sufficient for a reliable determination of 
$\mathcal{F}_{0}\lp R\rp$ slightly further away from the usual geometric 
touching configuration $R_0=9.06$ fm. To fix the parameters of the 
residual interaction (\ref{resin}), we first set the values $x_{c}=0$ 
and determine $\ovzer$ and $b$ which best reproduce the initial fields 
from the corresponding sets of eigenfunctions for protons and neutrons. 
One can then iterate until convergence using a gradual merging of 
previous $V^{\mathrm{(old)}}$ and current $V^{\mathrm{(calc)}}$ 
calculated potentials of the form
\bea
V^{\mathrm{(new)}}\lp r\rp=(1-\xi)V^{\mathrm{(old)}}\lp r\rp+
\xi V^{\mathrm{(calc)}}\lp r\rp
\eea
where $\xi$ starts from a small positive value and gradually approaches 
unity as we near convergence. The BCS calculation is carried out at each
step to provide the new occupation amplitudes. As input we used the
experimental proton and neutron pairing gaps. The result can be seen in
Fig. \ref{fig1}, depicted by the triple dot--dashed lines. This method 
generates a slightly compressed nucleus and is unable to create any 
significant clustering relative to experimental observations, so 
therefore we must turn to the surface term. The length parameter has 
the common value $B=1~\mathrm{fm}$ which can be inferred from the 
requirement of a narrow correction that does not disturb the low sp 
levels too much. The radial parameter should be slightly different for 
protons and neutrons in order for the surface corrections to overlap. 
One can then reproduce the decay width at a suitable radius using a 
common value of $x_{c}$, but this results in a final density of protons 
that is slightly too wide relative to the neutron density. Therefore, 
two values are tweaked around the common point, until the decay width is
reproduced optimally. The resulting fields are depicted in Fig. 
\ref{fig1} by continuous lines. The final set of values is shown in 
Table \ref{tab1}. These self--consistent fields are very similar to 
those phenomenologically introduced by \cite{Del13} when correcting the 
starting MF with a residual surface cluster. Notice the lowering of the
Coulomb barrier which in turn increases the decay width, as required in
order to reproduce the experimental value. In Fig. \ref{fig2} we use 
the same convention to represent the proton and neutron densities of the
parent nucleus. The dot--dashed lines represent the starting MF 
configuration of nucleons. The triple dot--dashed lines show the SCF 
density without surface correction terms in the residual interaction and 
the continuous lines represent the SCF density with surface correction 
appropriate for the experimentally observed decay width. One notices 
that this final configuration corresponds to an enhancement of the 
density near the surface. This is more evident for protons due to 
Coulomb repulsion, while neutrons have a broader, smoothly decreasing 
spatial extension. The resulting neutron skin thickness of 
$\ce{^{216}_{}Rn}$ is $\delta r_{np}=0.34~\mathrm{fm}$, a value 
significantly smaller than the $0.55$ fm determined from the MF 
approximation.

\begin{figure}
\centering 
\includegraphics[width=0.4\textwidth]{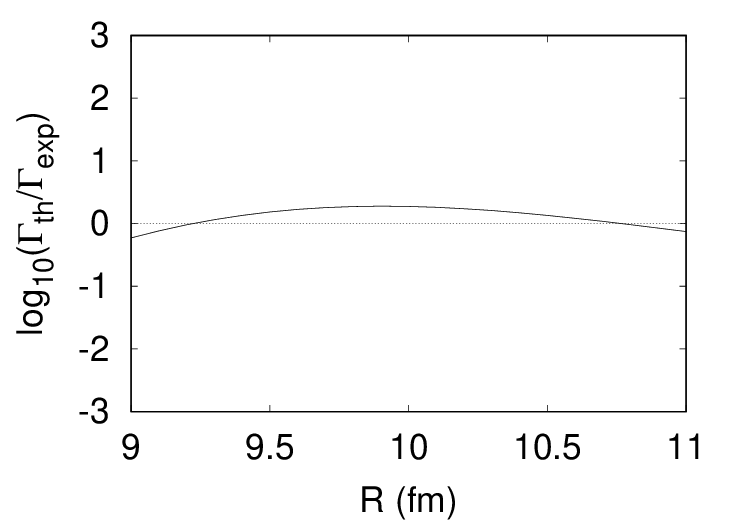}	
\caption{Logarithm of the ratio between the theoretical and experimental
decay width versus radius for the decay 
$\ce{^{216}_{}Rn}\rightarrow\ce{^{212}_{}Po}+\alpha$ calculated 
self--consistently from the residual interaction (\ref{resin}) 
parametrized as in Table \ref{tab1}.} 
\label{fig3}
\end{figure}

In a previous work \citep{Dum23} we discussed the extent to which 
decay widths calculated in the manner summarized in Subsection 
\ref{forma}  satisfy the plateau condition, namely that the computed 
value is independent of radius beyond the nuclear surface. Here, we 
managed to establish an approximate plateau condition at just a little 
under $10$ fm, as shown in Fig. \ref{fig3}. It amounts to slightly less 
than $1$ fm beyond the geometrical touching configuration. This radius 
lies beyond the Mott point, i.e. the value of the density has decreased 
to less than $10\%$ of the central value, which is the predicted 
threshold for the phase transition to an $\alpha$--condensate for 
symmetrical nuclear matter \citep{Rop98}. We stress that the values of 
$x_{c}$ were determined from an averaging condition
\bea
\langle\log_{10}\frac{\Gamma_{\mathrm{th}}\lp R\rp}
{\Gamma_{\mathrm{exp}}}\rangle=0
\eea
where the mean was considered over a range of $\pm 1$ fm around the 
peak. The averaged decay width thus calculated reproduces the 
experimental value. The spectroscopic factor 
\bea
s_{\alpha}=\intd_{0}^{\infty}\vert R
\mathcal{F}_{0}\lp R\rp\vert^{2}\mathrm{d}R
\eea  
quantifies how much of the parent gs configuration amounts to a 
clustered structure, our result being slightly over $3\%$. This gives
the probability of the parent gs being configured as a daughter core 
plus $\alpha$--particle. 

\begin{table}
\begin{tabular}{c c c c c c} 
$\tau$ & $\ovzer$ (MeV) & b (fm) & B (fm) & $r_{0}$ (fm) & $x_{c}$ \\
\hline
p      & ~~41.713       & 1.3    & 1      & 1.26         & 11.3 \\
n      &  241.467       & 0.6    & 1      & 1.30         & 16.7 \\
\hline
\end{tabular}
\caption{parameters of the residual interaction of Eq. (\ref{resin})
reproducing in a SCF the $\alpha$--decay width for the process 
$\ce{^{216}_{}Rn}\rightarrow\ce{^{212}_{}Po}+\alpha$.
}
\label{tab1}
\end{table}


\section{Summary and conclusions}
\label{conc}
We have developed a CHF SCF procedure starting from a WS MF and a 
residual SGI of the Wigner type parametrized from decay data. It is 
important to stress on the fact that the use of a surface term 
restoring the nuclear radius simplifies the standard cranked HFB method 
involving the generalized quasiparticle representation. We have shown 
that in this  manner clustering phenomena and particle emission can be 
calculated self--consistently, both the decay width and nuclear gs 
properties being simultaneously reproduced fairly well. We plan to 
perform an extension of this microscopic formalism to deformed nuclei,
in order to investigate the role played by $\alpha$-clustering on
$\alpha$, heavy cluster, $\beta^{\pm}$ and electromagnetic transitions.
A systematic investigation will allow us to parametrize the model and
provide predictions for cases of experimental interest, such as 
superheavy or exotic nuclei.

\section*{Acknowledgements}
This work  was supported by a grant of the Romanian Ministry of 
Education and Research PN 23 21 01 01/2023. Discussions with
Prof. Chong Qi (Stockholm) are gratefully acknowledged.




\bibliographystyle{elsarticle-harv} 
\bibliography{nuclear}






\end{document}